\documentclass[letterpaper]{jpconf}
\usepackage{graphicx}
\usepackage{amsmath}
\usepackage{subfig}
\begin{document}

\title{Results from the Neutral Current Detector phase of the Sudbury Neutrino Observatory}

\author{Ryan Martin for the SNO Collaboration}

\address{Department of Physics, Engineering Physics and Astronomy, Queen's University, 99 University Avenue, Kingston, Ontario, K7L 3N6, Canada}
\ead{rmartin@owl.phy.queensu.ca}

\begin{abstract}The Sudbury Neutrino Observatory (SNO) was a heavy water Cerenkov detector designed to solve the long-standing ``solar neutrino problem''; a discrepancy between the measured and predicted flux of electron-flavour solar neutrinos. SNO measured the rate of charged-current and neutral-current reactions of neutrinos in heavy water and was able to demonstrate that neutrinos from the Sun, produced in the electron flavour eigenstate, undergo flavour change on their way to the Earth, thus resolving the solar neutrino problem. The experiment was conducted in three phases, differing by the method for measuring the neutral current rate. This short paper summarizes results from the third phase of the experiment, which used an array of 36 strings of proportional counters filled with $^3$He to detect neutrons from the neutral-current reaction. When the data from the three phases is combined with solar and the KamLAND neutrino oscillation experiments, the resulting limits on the solar neutrino mixing angle and mass-squared difference are $\theta_{12} = 34.4^{+1.3}_{-1.2}$ degrees and $\Delta m_{12}^2 =7.59^{+0.19}_{-0.21}\times 10^{-5}$eV$^2$, respectively.
\end{abstract}

\section{Introduction}
The Sudbury Neutrino Observatory \cite{snonim} was located in the Vale-Inco mine in Sudbury, Ontario, Canada under two kilometers of rock overburden ($\sim$6000\,m of water equivalent). The detector was comprised of a 12\,m diameter acrylic vessel filled with 1000 tonnes of ultra-pure heavy water (D$_2$O) monitored by $\sim$9500 8'' photo-multiplier tubes mounted on an 18\,m diameter support structure. SNO detected Cerenkov electrons induced by $^8$B solar neutrinos \cite{ssm1} \cite{ssm2} through three different reactions on deuterium (d), CC, NC and ES:
\begin{align*}
&\nu_e + d \rightarrow p + p +e^- &(CC)\nonumber\\
&\nu_x + d  \rightarrow p + n + \nu_x &(NC)\nonumber\\
&\nu_x +e^-\rightarrow \nu_x +e^- &(ES)
\end{align*}

The experiment was conducted in three phases, differing in the method that was used to detect the neutron from the NC reaction. In the first phase of the experiment \cite{d2o1} \cite{d2o2} \cite{d2o3}, Cerenkov electrons from the Compton scattering of 6.25\,MeV gamma-rays from neutron captures on deuterium were detected. In the second phase \cite{salt1} \cite{salt2}, 2 tonnes of salt (NaCl) were dissolved in the heavy water to benefit from the enhanced detection efficiency of neutron captures on $^{35}$Cl. Finally, in the third phase of the experiment \cite{ncd1} \cite{ncdnim}, 36 strings of proportional counters filled with $^3$He were deployed in the heavy water to detect thermal neutrons in an independent way using the reaction:
\begin{align*}
n+^3\!He\rightarrow p + ^3\!H + 764\,keV
\end{align*}
Four additional counters, filled with $^4$He, insensitive to neutrons, were also deployed to measure backgrounds.

The first two phases of the experiment confirmed the standard solar model prediction for the flux of neutrinos from the $^8$B decay. Within the framework of neutrino oscillations and the MSW effect \cite{msw}, one can combine the data from SNO with other solar neutrinos experiments and the KamLAND \cite{kamland} reactor anti-neutrino experiment to determine limits on the solar neutrino mixing angles. Using the first two phases of SNO as well as the other experiments \cite{salt2}, one obtains $\theta_{12} = 33.9^{+1.6}_{-1.6}$ degrees and $\Delta m_{12}^2 =8.00^{+0.4}_{-0.3}\times 10^{-5}$eV$^2$,

\section{The Neutral Current Detectors}
The proportional counters measured between 2\,m and 3\,m in length and were combined into strings measuring between 9\,m and 11\,m. The counters had a 5\,cm diameter and were made of ultra-pure nickel cylinders fabricated by chemical vapor deposition \cite{ncdnim} with a copper anode wire. These Neutral Current Detectors (NCD) were made very pure to limit background events (primarily alpha emitters from the uranium and thorium decay chains).

Each NCD event was recorded independently by two different channels. One channel recorded the charge of the events while the other collected scope traces of the charge on the anode as a function of time. The charge of the events is proportional to the energy deposited in the counters, whereas the scope trace allowed for rejection of non-physical events. The scope traces were recorded through a logarithmic amplifier to extend the dynamic range.

\section{Analysis of the NCD data}
The number of neutrons in the counters was then determined  with a statistical analysis of the recorded charge (energy). The shape of the charge spectrum for neutrons was obtained and verified with several calibration data. In particular, a $^{24}$Na solution was dissolved in the heavy water; this produced a source of uniform photo-disintegration neutrons. The neutron calibration data was also used to measure the neutron detection efficiency.

Alpha events in the energy spectrum where modeled using Monte-Carlo simulations of the main alpha emitting backgrounds. The Monte-Carlo simulation was verified by comparison with data from the NCD strings that were filled with $^4$He (insensitive to neutrons). Finally, two of the strings showed evidence for instrumental events; distributions for these events were obtained from the strings and included in the final statistical analysis, although the data from those strings was not kept. Four more strings were rejected from the final analysis which was carried out on the remaining 30 strings.

The data from the NCD strings was combined with the data from the photo-multiplier tubes into one likelihood function with floating neutrino fluxes and systematic uncertainties. Due to the large number of parameters in the likelihood function, a Bayesian approach was adopted and a Markov-Chain Monte-Carlo \cite{mcmc} was used to sample the likelihood function. The parameters (neutrino fluxes, systematic parameters) and their uncertainties were then determined by fitting normal distributions to the posteriors returned from the Markov-Chain Monte-Carlo.

\begin{figure}[h]
  \centering
\includegraphics[width=0.5\textwidth]{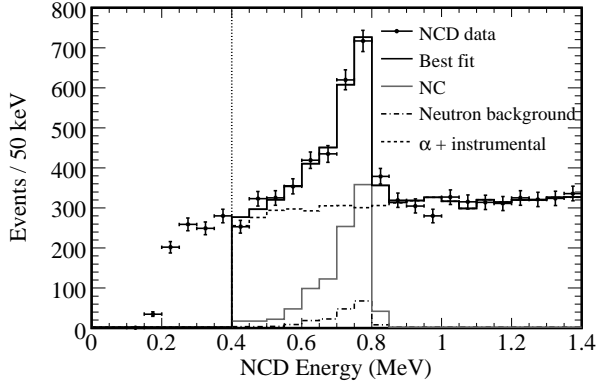}
\caption{\label{EnergyFit}Fit to the NCD energy spectrum. The neutron energy distribution was obtained from calibration data, the alpha emitters were modeled with a Monte-Carlo simulation and the instrumentals were obtained from two of the NCD strings that showed evidence for events other than neutrons or alphas. Figure reproduced from \cite{ncd1}.}
\end{figure}

\begin{figure}[ht]
   \centering
   \subfloat{\includegraphics[width=0.7\textwidth]{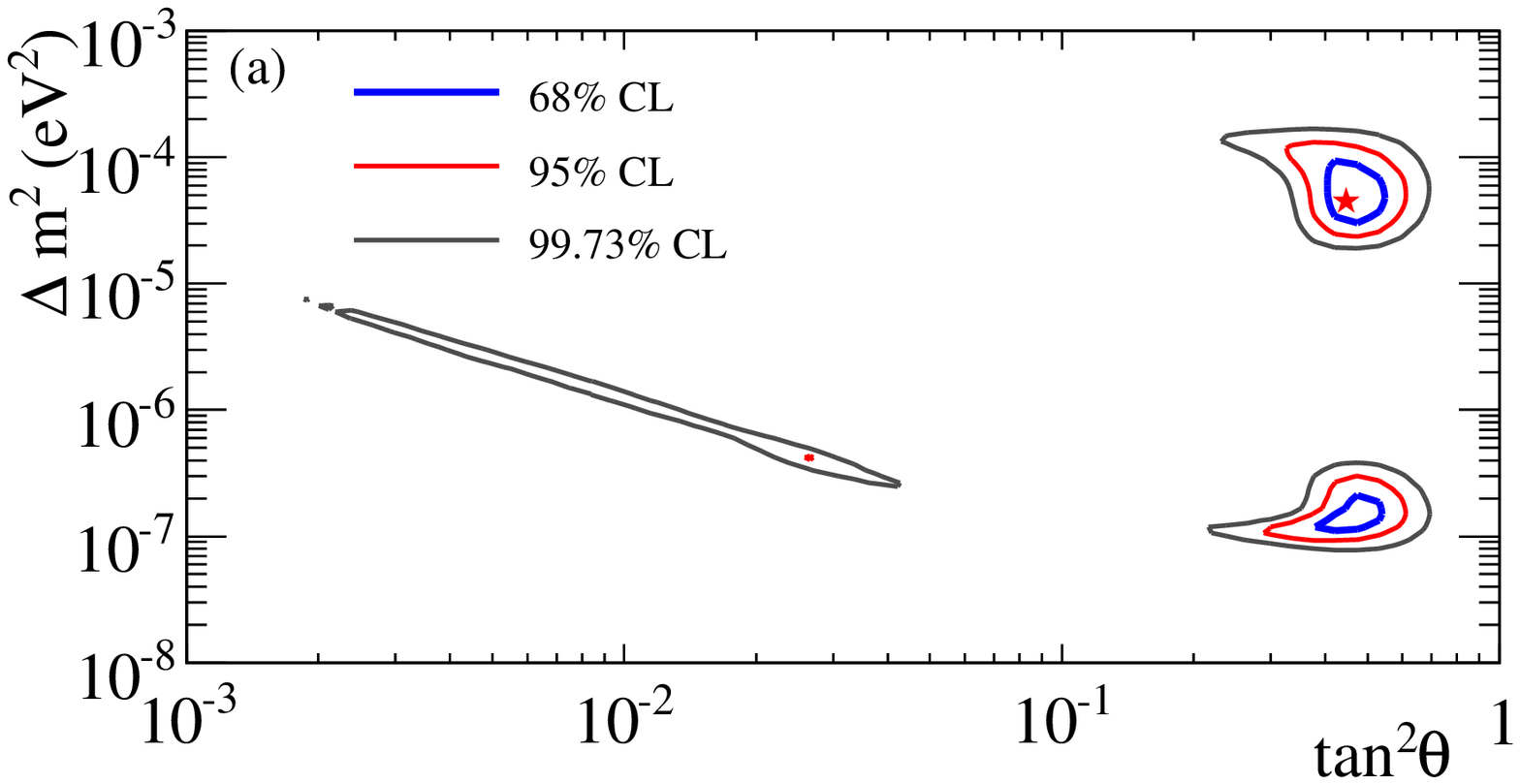}}
   \subfloat{\includegraphics[width=0.7\textwidth]{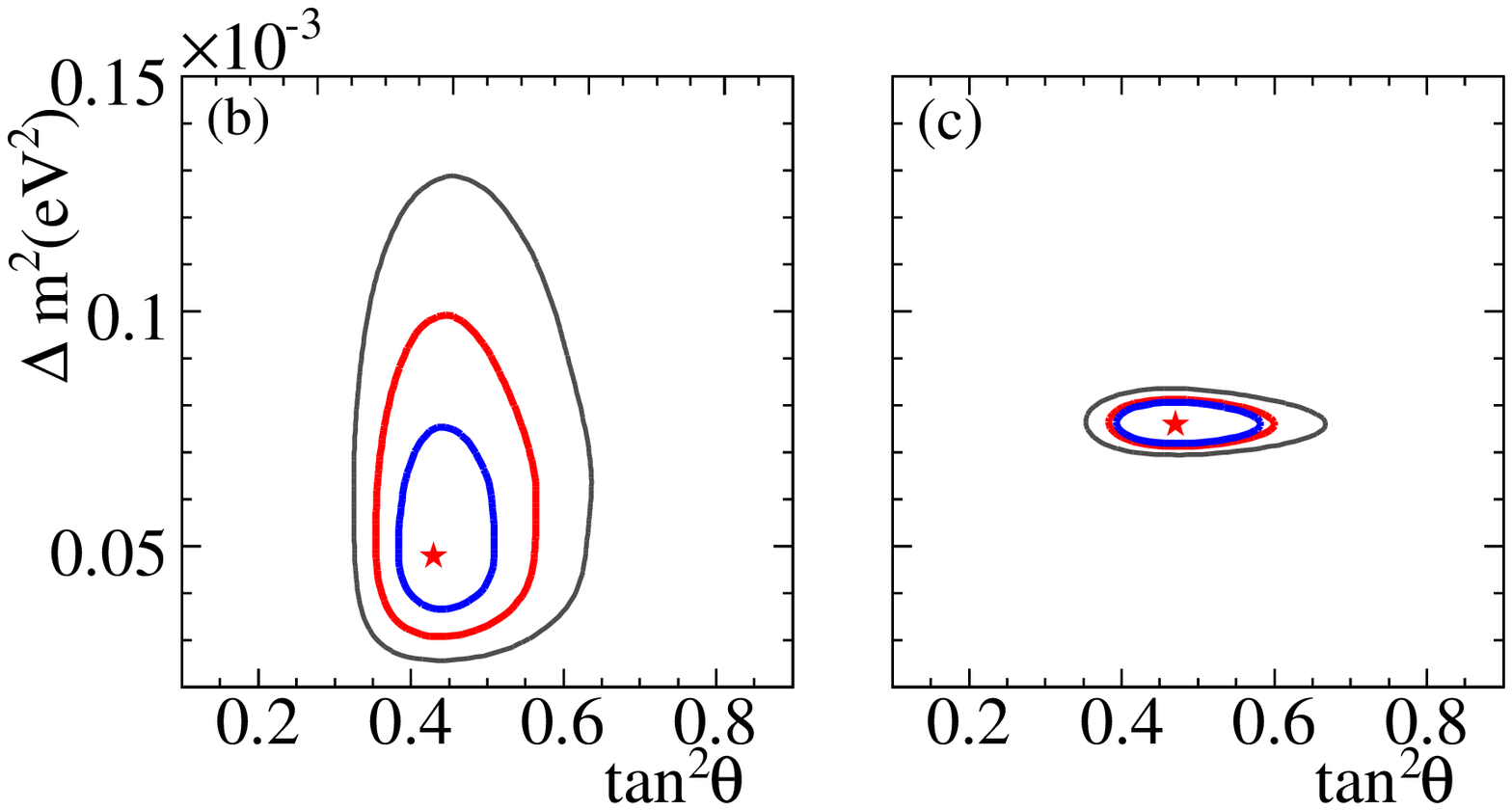}}
\caption{\label{contours}Allowed ranges for solar neutrino mixing parameters using all SNO Data (a), all solar experiments (SuperK, Cl, Ga and Borexino) (b) and combined with data from KamLAND (c). The resulting mixing parameters are determined to be $\theta_{12} = 34.4^{+1.3}_{-1.2}$ degrees and $\Delta m_{12}^2 =7.59^{+0.19}_{-0.21}\times 10^{-5}$eV$^2$. Figure reproduced from \cite{ncd1}}
\end{figure}

Figure \ref{EnergyFit} shows the resulting fit of the data in NCD energy. The data from the NCD phase of the experiment was then used to restrict the limits on the neutrino mixing angles and results in the contours shown in Figure \ref{contours}. When data from all solar neutrino experiments is combined with KamLAND data, one obtains $\theta_{12} = 34.4^{+1.3}_{-1.2}$ degrees and $\Delta m_{12}^2 =7.59^{+0.19}_{-0.21}\times 10^{-5}$eV$^2$ \cite{ncd1}.

\newpage 

\end{document}